\documentclass[11pt]{article}
\usepackage[margin=1in]{geometry}
\usepackage{graphicx}
\usepackage{booktabs}
\usepackage{hyperref}
\usepackage{enumitem}
\usepackage{amsmath}
\usepackage{amsfonts}
\usepackage{amssymb}
\usepackage{float}
\usepackage{caption}
\hypersetup{
    colorlinks=true,
    urlcolor=blue,
}
\usepackage{fancyhdr}
\title{Using the NANDA Index Architecture in Practice: An Enterprise Perspective}
\author{Sichao Wang (Cisco Systems), \and Ramesh Raskar (MIT), \and Mahesh Lambe (Unify Dynamics), \and Pradyumna Chari (MIT), \and Rekha Singhal (Tata Consultancy Services), \and Shailja Gupta (CMU), \and Rajesh Ranjan (CMU), \and Ken Huang (Distributedapps.AI)}
\date{}

\begin{document}

\maketitle

\section*{Abstract}
The proliferation of autonomous AI agents represents a paradigmatic shift from traditional web architectures toward collaborative intelligent systems requiring sophisticated mechanisms for discovery, authentication, capability verification, and secure collaboration across heterogeneous protocol environments. This paper presents a comprehensive framework addressing the fundamental infrastructure requirements for secure, trustworthy, and interoperable AI agent ecosystems.\\

We introduce the NANDA (Networked AI Agents in a Decentralized Architecture) framework, providing global agent discovery, cryptographically verifiable capability attestation through AgentFacts, and cross-protocol interoperability across Anthropic's Modal Context Protocol (MCP), Google's Agent-to-Agent (A2A), Microsoft's NLWeb, and standard HTTPS communications. NANDA implements Zero Trust Agentic Access (ZTAA) principles, extending traditional Zero Trust Network Access (ZTNA) to address autonomous agent security challenges including capability spoofing, impersonation attacks, and sensitive data leakage.\\

The framework defines Agent Visibility and Control (AVC) mechanisms enabling enterprise governance while maintaining operational autonomy and regulatory compliance. Our approach transforms isolated AI agents into an interconnected ecosystem of verifiable, trustworthy intelligent services, establishing foundational infrastructure for large-scale autonomous agent deployment across enterprise and consumer environments.\\

This work addresses the critical gap between current AI agent capabilities and infrastructure requirements for secure, scalable, multi-agent collaboration, positioning the foundation for next-generation autonomous intelligent systems.

\section{Introduction}
The emergence of autonomous AI agents marks a fundamental transformation in computational architecture, shifting from traditional client-server paradigms toward dynamic, collaborative intelligent systems capable of independent reasoning, decision-making, and cross-platform interaction.\\

Contemporary AI agents possess sophisticated capabilities that enable them to autonomously discover resources, negotiate collaborations, and execute complex tasks across distributed environments without continuous human oversight. This evolutionary leap necessitates a corresponding advancement in underlying infrastructure to support secure, trustworthy, and scalable agent-to-agent interactions. \\

Traditional web application architectures rely upon well-established infrastructural foundations including domain name resolution (DNS) infrastructure, public key infrastructure and TLS protocol that provides secure communication channels over the Internet. TLS ensures data confidentiality and integrity between two communicating endpoints, such as a web browser and a web server. However, the autonomous nature of AI agents introduces unprecedented challenges that transcend the capabilities of existing web infrastructure frameworks. \\

AI agents operate as autonomous entities with distinct life cycles, claimed capabilities, and decision-making authority, requiring sophisticated mechanisms for mutual discovery, authentication, capability verification, and trust establishment. Unlike traditional web services that respond predictably to predefined requests, AI agents must dynamically assess the trustworthiness and capabilities of potential collaborators while protecting themselves against sophisticated attack vectors including capability spoofing, impersonation, and supply-chain poisoning. The autonomous nature of these systems amplifies both the potential benefits and risks associated with inter-agent collaboration. \\

The current landscape of AI agent frameworks demonstrates significant fragmentation, with multiple incompatible protocols including Anthropic's Modal Context Protocol (MCP)~\cite{anthropic_mcp_2024}, Google's Agent-to-Agent (A2A) framework~\cite{google_a2a_2025}, Microsoft's NLWeb~\cite{microsoft_nlweb_2025}, and various proprietary implementations. This protocol heterogeneity creates isolated ecosystems that prevent seamless collaboration between agents developed using different technological foundations, limiting the potential for comprehensive multi-agent solutions. \\


This paper addresses these fundamental challenges to enterprise use cases of the agentic web, through the lens of the NANDA (Networked AI Agents in a Decentralized Architecture) framework~\cite{raskar2025beyond}, a comprehensive infrastructure solution designed to enable secure, verifiable, and interoperable AI agent ecosystems. The NANDA architecture provides cryptographically verifiable agent discovery through AgentFacts metadata, cross-protocol interoperability through standardized adapter mechanisms, and robust security frameworks implementing Zero Trust Agentic Access (ZTAA) principles specifically designed for autonomous agent environments. \\

\subsection{Common Terminology}

\begin{itemize}
    \item \textbf{Autonomous AI Agent or Agent:} Software entity with specific goal-directed reasoning, memory, and the ability to initiate actions, migrate, or spawn helpers without continuous human supervision.
    \item \textbf{Agentic AI:} refers to the broader concept of AI systems that can autonomously make decisions and take actions, often involving multiple interacting AI agents working together to achieve complex goals.
    \item \textbf{AI Agent Index/Registry:} Authoritative system (central, decentralized, or hybrid) that stores cryptographic identifiers, capability descriptors, trust metadata, and audit logs for agents.
    \item \textbf{NANDA Index:} A lightweight index/registry system for autonomous agents~\cite{raskar2025beyond}. Provides a Web interface for browsing agents and a REST API for programmatic access. The system supports Google OAuth authentication and implements full CRUD operations with ownership controls. 

    
    \item \textbf{Verifiable Metadata (AgentFacts):} All meta-data that define the identities (cryptographic keys) of AI agents that allow other agents to verify who the agent is, what capabilities it can offer.

\end{itemize}

\section{A Look Back: Domain Registration in the Traditional Web}
Historically, and continuing to the present day, establishing website for any purpose requires a fundamental two-step process: domain name registration followed by obtaining a public certificate from a Certificate Authority, especially for secure, public-facing sites.\newline 

Domain registration involves securing a unique web address through an accredited registrar (such as GoDaddy, Google, or NameCheap), enabling any Internet user to access the website via a human-readable web address. To establish secure HTTPS connections, the server must also present a valid TLS certificate issued by a recognized CA. This seemingly straightforward interaction relies upon a sophisticated underlying infrastructure encompassing DNS resolution~\cite{mockapetris1987dns}, TCP handshakes, TLS sessions, and HTTP(S) message exchange to facilitate seamless browser-to-website communication~\cite{rescorla2000https}. \\

DNS translates the domain name into the corresponding IP address, enabling connection initiation. In many cases—particularly for Content Delivery Networks (CDNs) and globally distributed services—Anycast routing is used to direct traffic to the nearest server or data center, optimizing latency and reliability. The TLS certificate ensures encrypted communication over TCP, protecting data integrity and establishing trust between the client and server.\\

The realm of AI agents presents analogous challenges—discovery, identification, authentication, trust establishment, and communication—while requiring novel approaches tailored to their unique characteristics. Unlike traditional web services, AI agents possess autonomous capabilities and must account for both claimed and actual functional capabilities, introducing a distinctive yet crucial consideration in agent design and deployment.\newline

As illustrated in Figure 1, establishing an AI agent and properly configuring it for seamless collaboration with other Internet-based agents involves several critical implementation steps.\newline

From an architectural standpoint, a fundamental distinction between classical web applications and AI agents lies in the latter's reliance on microservices architectures. This design paradigm enables modularity and independent scalability of individual AI agent components during development, deployment, and operational scaling. When an AI agent requires additional resources—such as a compute-intensive recommendation engine—these components can be scaled independently without impacting other system elements. Furthermore, microservice architecture inherently enhances agent autonomy and enables event-driven operational models.\newline



\begin{figure}[ht]
    \centering
    \includegraphics[width=\textwidth]{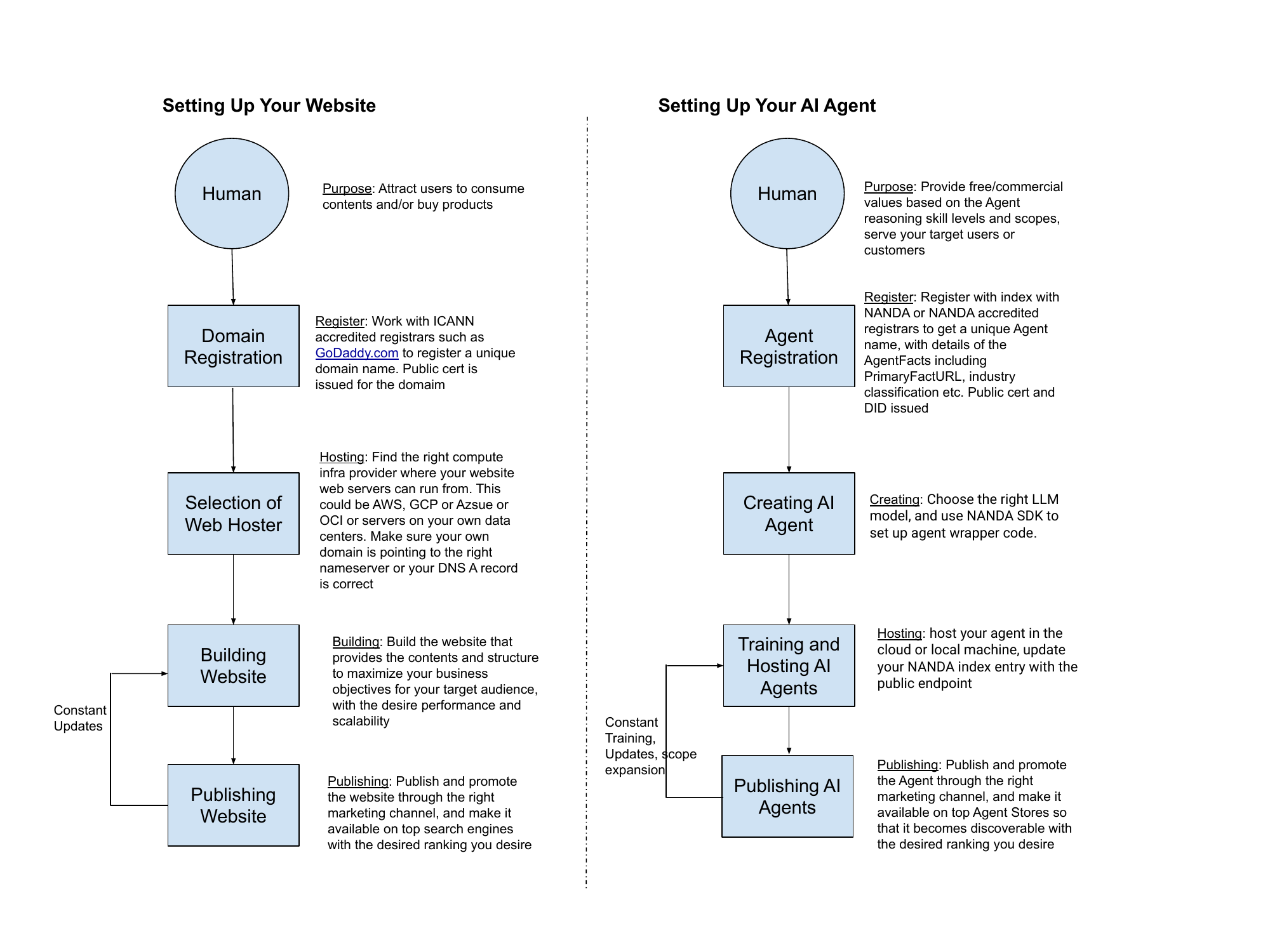}
    \caption{Building and Running an AI Agent}
    \label{fig:1}
\end{figure}

\section{The Setting: Agentic Web and the NANDA Index}
To conceptualize AI agents within a computational framework, we propose considering them as virtual entities that exhibit human-like behavioral characteristics: they possess defined purposes, maintain distinct life cycles, operate under assigned contracts from their human owners, and demonstrate cognitive reasoning capabilities that enable autonomous decision-making and action execution. This perspective provides valuable insight into the operational dynamics AI agents experience when working and collaborating within multi-agent environments.\newline

\begin{figure}[H]
    \centering
    \includegraphics[width=\textwidth]{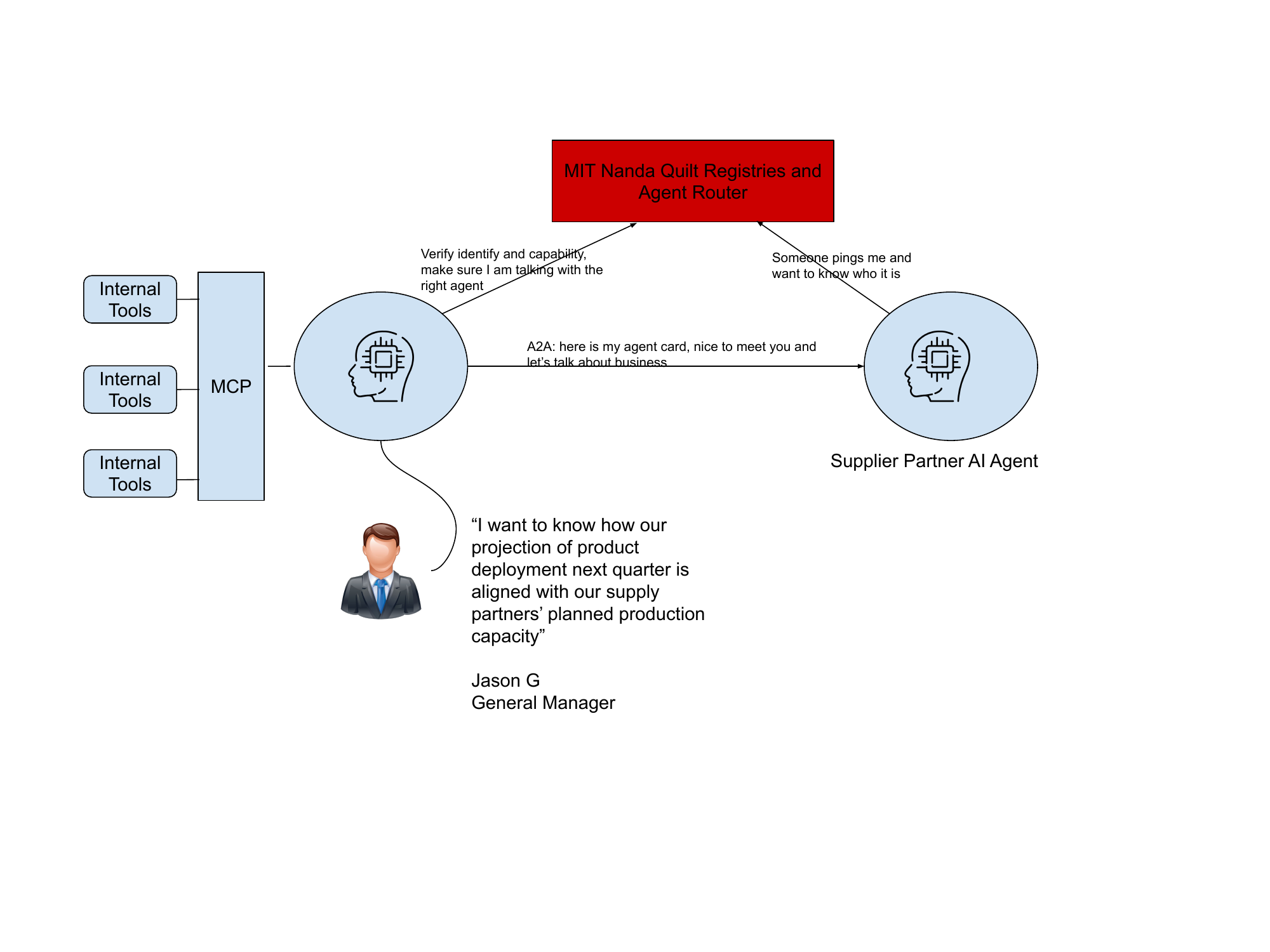}
    \setlength{\abovecaptionskip}{-5pt}
    \caption{Agentic Architecture Breakdown}
    \label{fig:2}
\end{figure}

The scenario illustrated in Figure 2 demonstrates how Jason interfaces with an AI application through a Large Language Model (LLM), issuing instructions and defining tasks for offline or on-demand execution.\newline

Within the traditional Internet ecosystem, users primarily employ browsers to access websites for content consumption and interactive functionality, such as account creation and financial transactions through button clicks and data entry. Web applications in this paradigm are characterized by their relatively static nature, focusing on HTTP(S) request handling from users and browsers, establishing trusted sessions with client browsers, and constructing responses through database access and business logic processing.\newline

AI agents, as depicted in the aforementioned figure, operate at the foundational layer of the Agentic AI application stack, functioning behind the scenes to accomplish user-defined objectives through interactions with other agents, both internal and external to their operational environment. Agentic AI systems demonstrate enhanced autonomy compared to traditional web applications, leveraging collaborative relationships with trusted internal and external agents to fulfill their designated functional responsibilities.\newline





\subsection{Interoperability and Trust}
Within enterprise environments, organization-owned AI agents establish communication pathways with internal and external tools through the Modal Context Protocol (MCP), enabling individual AI agents or Large Language Models to securely and efficiently access and utilize external tools and data sources. Additionally, these agents leverage the Agent-to-Agent (A2A) protocol for horizontal integration, facilitating communication and collaboration among disparate, independent agents while orchestrating workflows across diverse systems and domains.\newline

The NANDA Index provides an extensible open architecture that enables agents to perform sophisticated filtering operations based on AgentFacts data, ensuring target agents meet trust and safety requirements for collaborative engagement. The NANDA Index's open architecture supports third-party solution integration to enrich AgentFacts, provide independent capability classification ratings (spanning domains such as financial consulting, travel booking, and education), and generate reputation scores. These supplementary third-party attributes maintain certifiable and verifiable characteristics.\newline

AgentFacts are conceptualized as verifiable claims requiring W3C Verifiable Credential v2 cryptographic attestation \cite{nanda-why}, with credential issuers encompassing enterprises, consortiums, and federated certification authorities operating within domain-specific trust zones capable of mutual cross-signing. Each claim maintains linkage to credentialing paths anchored in issuer Decentralized Identifiers (DIDs) with revocation status management via VC-Status-List mechanisms. This framework necessitates several co-dependencies: W3C Verifiable Credential infrastructure, DID resolution systems, cross-signing protocols between trust zones, and VC-Status-List revocation mechanisms.\newline

NANDA has developed a comprehensive infrastructure utilizing verifiable claims to prevent AI agents from engaging in capability spoofing, impersonating reputable actors, or conducting supply-chain poisoning, Sybil attacks, and traffic diversion attacks.\newline

Furthermore, the NANDA Index operates as a global discoverability layer where agents across heterogeneous protocols—including Anthropic's MCP, Google's A2A, Microsoft's NLWeb, and standard HTTPS—can register and discover each other across all existing protocol registries. The NANDA Adapter subsequently establishes actual communication channels, automatically configuring protocol translations to enable seamless cross-framework agent communication.\newline

This architectural synergy enables comprehensive cross-platform agent interoperability: an MCP-based assistant can directly interface with an A2A inventory system or an NLWeb calendar service, establishing a unified ecosystem where agents collaborate seamlessly regardless of their underlying technology stack, while preserving organizational autonomy over individual infrastructure components.\newline

\section{Use Cases}
\subsection{Agentic AI System Component Stack}

The following analysis presents a comparative examination between traditional user/browser interactions with conventional web application architectures and AI agent interactions within the emerging agentic framework paradigm.\newline

\begin{figure}[H]
    \centering
    \includegraphics[width=\textwidth]{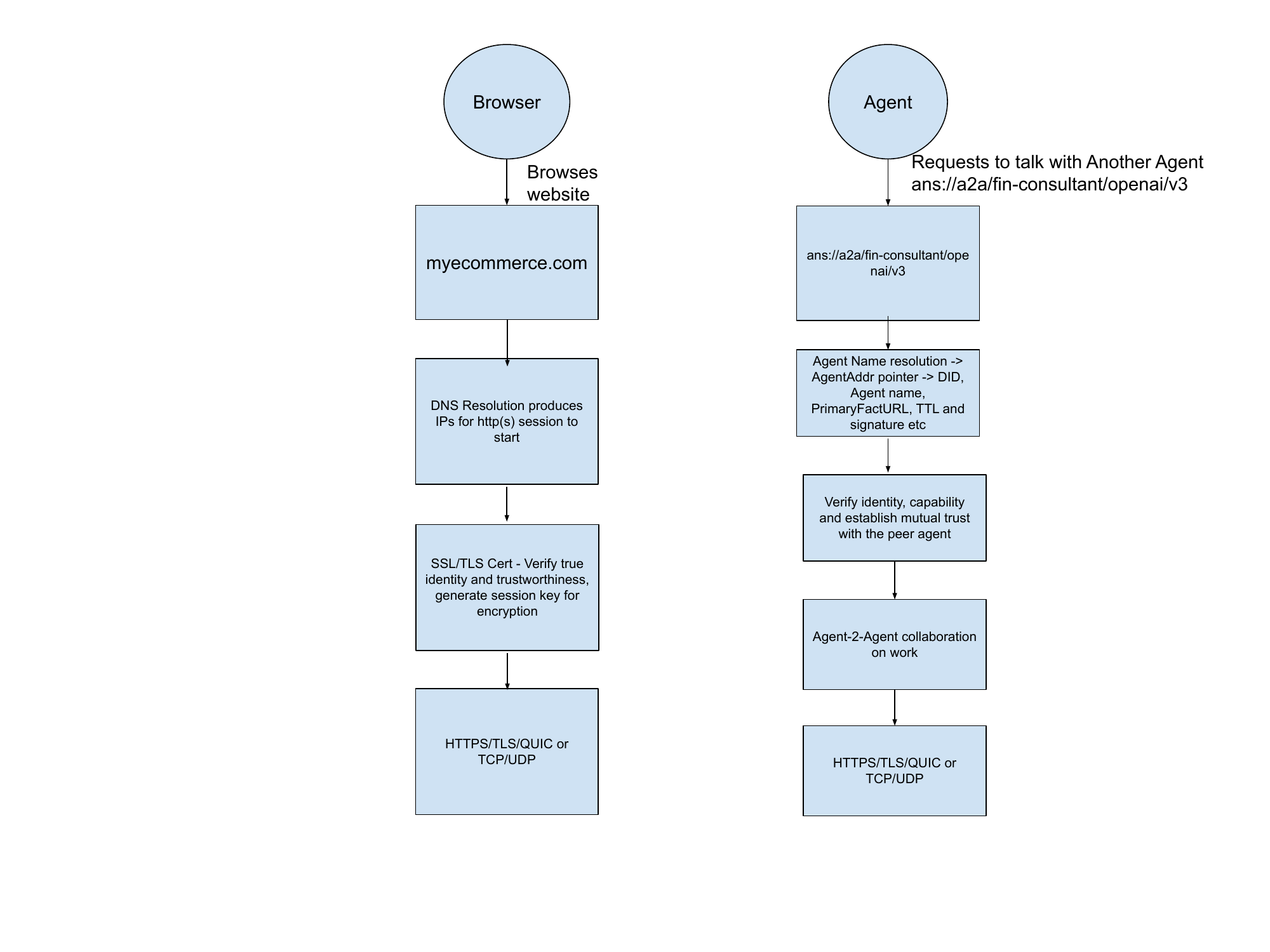}
    \setlength{\abovecaptionskip}{-5pt}
    \caption{Browser-to-Webserver vs Agent-to-Agent Comparison}
    \label{fig:3}
\end{figure}

The fundamental distinction within the Agentic AI workflow lies in the requirement for AI agents to perform comprehensive verification procedures before establishing collaborative relationships. Specifically, AI agents must verify target agent identities, execute name-to-agent address resolution protocols, and leverage AgentFacts data to validate agent capabilities while implementing filtering mechanisms based on risk assessment and reputation attributes.\newline



\subsection{An Enterprise AI Architecture}

The following presents an exemplary enterprise architecture that leverages AI agents to construct an AI Assistant Application, optimizing and automating sales and marketing operations under human employee supervision.\newline

Consider a representative use case wherein sales executive Tom encounters operational inefficiencies within the sales process encompassing lead discovery, Proof-of-Concept (POC) preparation, deal registration, negotiations, closing procedures, and subsequent service delivery and deployment phases. His team confronts challenges stemming from fragmented processes and disparate toolsets including Customer Relationship Management (CRM) systems, inventory management platforms, financial pricing and discounting applications, marketing campaign email tools, video creation platforms, and product package and feature enablement management systems.\newline

\begin{figure}[H]
    \centering
    \includegraphics[width=\textwidth]{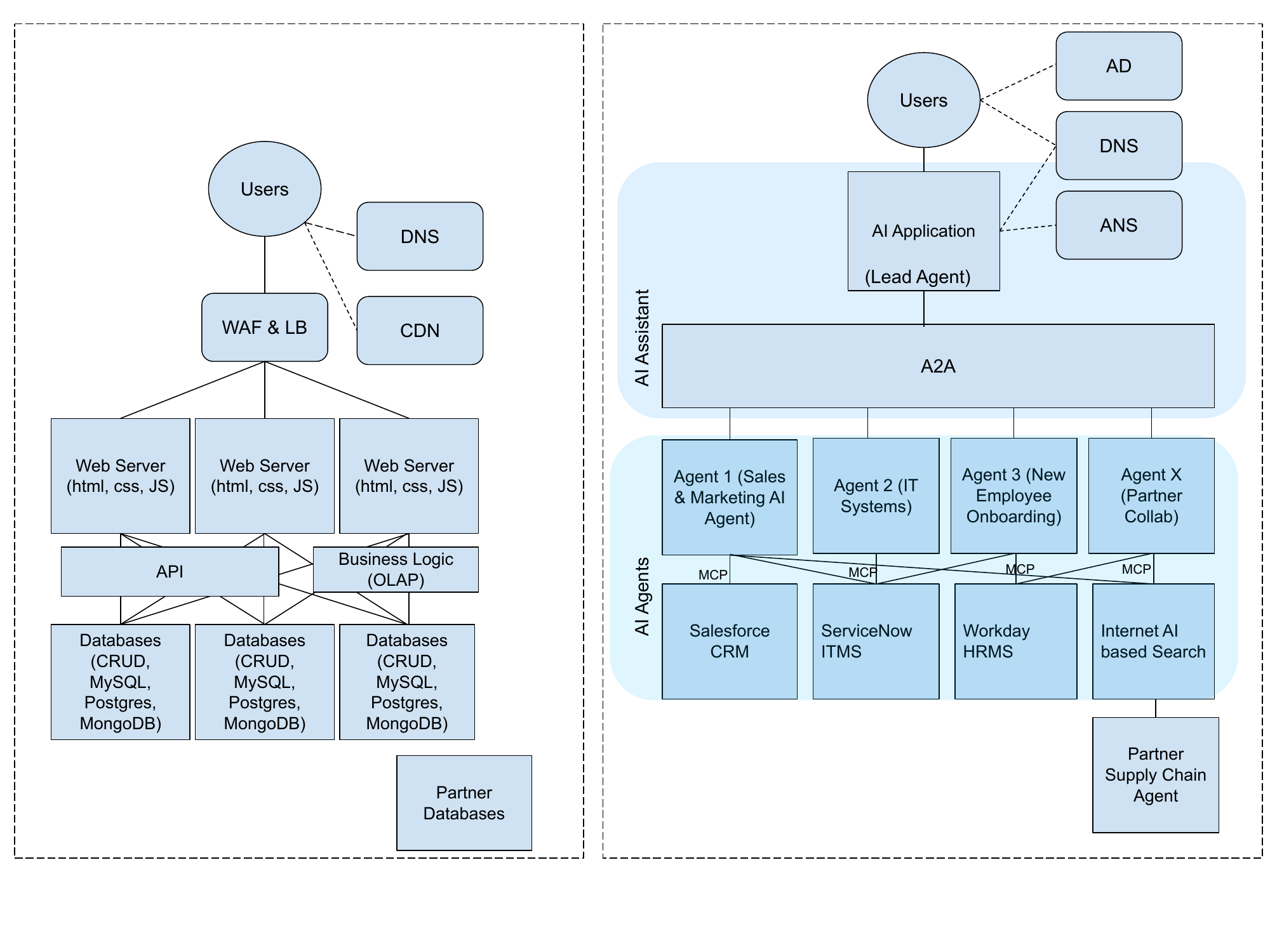}
     \setlength{\abovecaptionskip}{-5pt}
    \caption{An Enterprise Web Architecture vs Agentic Architecture}
    \label{fig:4}
\end{figure}

This operational complexity and associated overhead present significant challenges and costs when generating customer quotes based on desired feature sets or renewal quotes for existing customers.\newline

Tom seeks to address these inefficiencies by implementing a targeted solution that leverages AI agents to automate the notification process for revenue-impacting customers regarding new feature releases aligned with their requirements. This automated process should encompass feature descriptions, value propositions, and supporting collateral materials including instructional videos demonstrating configuration and usage procedures. The entire workflow requires optimization through an agentic AI solution.\newline

\begin{figure}[H]
    \centering
    \includegraphics[width=\textwidth]{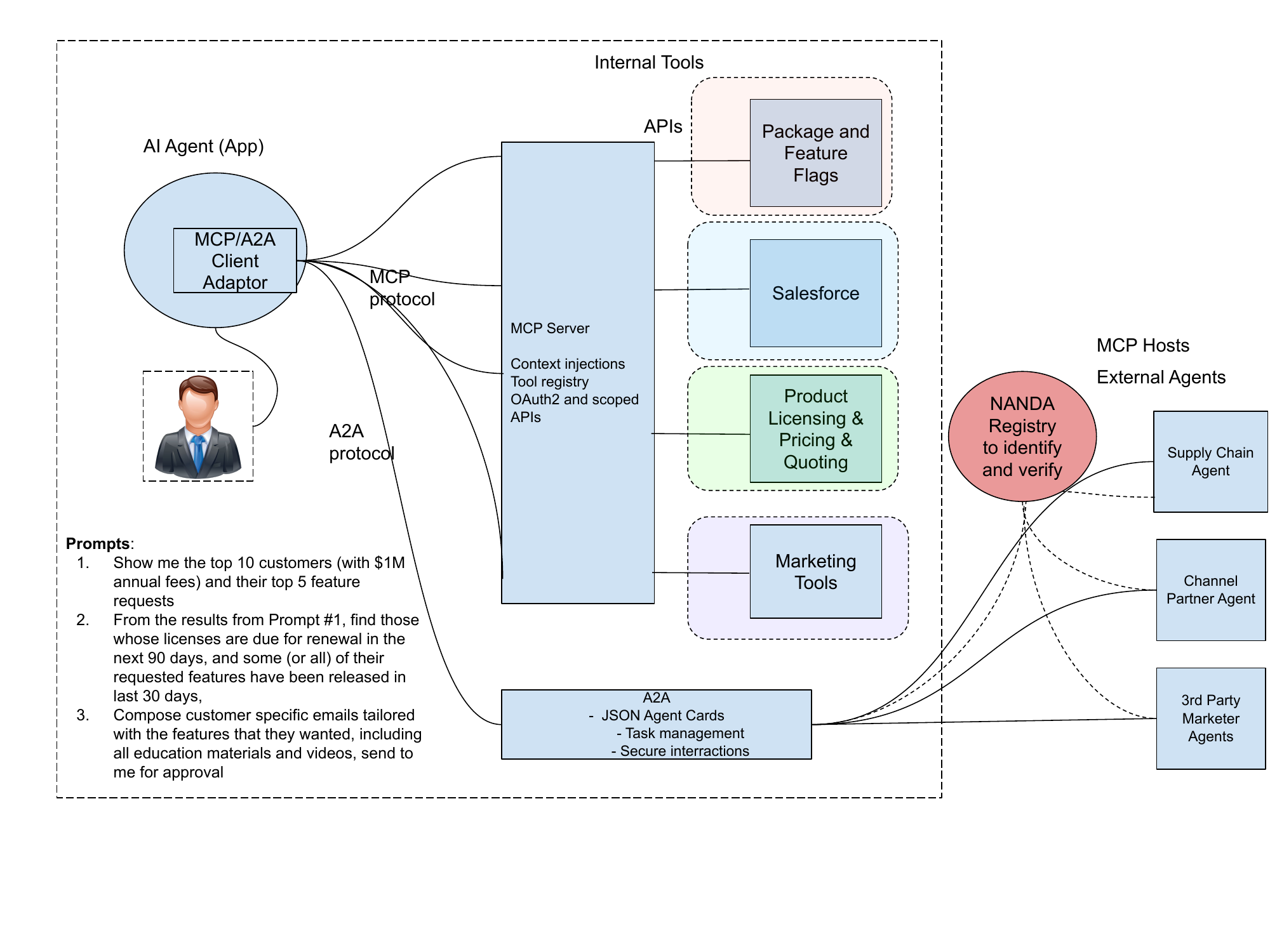}
    \setlength{\abovecaptionskip}{-5pt}
    \caption{An Enterprise Agentic Use Case for Product Operations}
    \label{fig:5}
\end{figure}

The contemporary business environment presents an opportunity for Sales and Marketing teams to realize substantial operational improvements through Agentic AI implementation. As illustrated in Figure 5, the Sales AI application functions as an orchestration layer constructed upon a network of internal AI agents, each representing specialized business systems including Aha! for feature management, Salesforce for customer relationship management, pricing and licensing tools for quote generation, and email marketing platforms. These agents interact through standardized inter-agent protocols and frameworks (such as MCP or A2A), enabling the AI application to retrieve pertinent information, perform reasoning operations, and construct responses utilizing an embedded Large Language Model (LLM) in response to human-generated prompts.\newline

The potential cost savings and efficiency improvements are substantial; however, these benefits depend upon the existence of a robust and secure foundational infrastructure. When an agent requires collaboration with external agents across the Internet, it must leverage the NANDA Registry to discover target agents, authenticate and verify their identities, and validate their claimed capabilities. The NANDA Index operates as a universal handshake layer that transforms isolated AI agents into an interconnected network of intelligent services. It functions as a global discovery layer where agents utilizing diverse protocols—including Anthropic's MCP, Google's A2A, Microsoft's NLWeb, and standard HTTPS—can register and discover each other across existing protocol registries. The NANDA Adapter establishes actual communication channels and automatically configures protocol translations to enable seamless cross-framework communication.\newline

\section{A Retrospective Look: Security on the Web}

For decades, web-based Software-as-a-Service (SaaS) platforms have been persistently challenged by server-side vulnerabilities, including SQL injection attacks (encompassing numerous SQL injection variants) that enable attackers to inject malicious SQL code into applications for unauthorized database access or modification, and Cross-Site Scripting (XSS) vulnerabilities that facilitate the injection of malicious scripts into websites, frequently resulting in the theft of sensitive personal data. The Open Web Application Security Project (OWASP) Top 10 continuously monitors and publishes findings regarding critical vulnerabilities that present exploitation opportunities for attackers seeking to compromise web security.\\

Table 1 presents a comprehensive analysis of the requirements for addressing security challenges across Large Language Model (LLM) architectures and AI applications, while simultaneously providing security protection for AI agents as they navigate Internet environments and collaborate with external entities to accomplish designated tasks:\\

\begin{itemize} 
\item \textbf{AI Model Security:} Model security encompasses protection against prompt injection attacks, supply chain vulnerabilities, sensitive data leakage, and data or model poisoning incidents.
\item \textbf{Agent Security:} Agent security focuses on ensuring AI agent compliance with IT administrator-defined security policies while preventing compromise or sensitive data leakage during Internet-based collaborative exploration activities.
\end{itemize}

\begin{table}[h]
\centering
\begin{tabular}{|p{3cm}|p{5.5cm}|p{5.5cm}|}
\hline
\textbf{Security Aspect} & \textbf{Web/Internet} & \textbf{Agentic AI} \\
\hline
\textbf{Discovering Targets} & SafeSearch & Agentic SafeSearch: requires verification through NANDA index and implements reputation-based capability classification filtering \\
\hline
\textbf{Communicating with Target Destinations} & URL filtering, application classification, content category specification within Acceptable Use Policy (AUP) & Enforcement of agent classification filtering based on IT security policies governing external AI agent association types and categories, data exchange parameters with target agents, and optional Data Loss Prevention (DLP) policy implementation on exchanged data \\
\hline
\textbf{Robust Security Hygiene} & Vulnerability scanning and patching, static and dynamic code analysis, input parameter validation, stored procedure implementation & Ensuring model training with clean, unpoisoned datasets and implementing defense mechanisms against direct and indirect prompt injection attacks \\
\hline
\textbf{Security Practice \& Protections} & Penetration testing, Web Application Firewall (WAF) deployment, identity posture management, two-factor authentication (e.g., zero-trust access defense such as Cisco Secure Access), Transport Layer Security (TLS) encryption & Review and enhancement of Modal Context Protocol (MCP) context sharing and inter-agent trust mechanisms, DLP enforcement, context validation and sanitization, robust authentication implementation, comprehensive monitoring and logging, and least-privilege context sharing protocols \\
\hline
\end{tabular}
\caption{Security Comparison: Web/Internet vs Agentic AI}
\end{table}

SafeSearch represents a filtering technique employed by search engines such as Google and Bing to exclude explicit or inappropriate content from search results. This mechanism is designed to establish safer online environments, particularly for families, educational institutions, and organizations managing search capabilities for others, by blocking images, videos, and websites containing potentially harmful material.\newline

A comparable solution is essential for AI agent protection, ensuring collaborative engagement exclusively with safe and trustworthy agents. Given that Agentic AI autonomy enables action execution without continuous human oversight, this capability assumes critical importance. An agent utilized by a children's educational AI system could potentially exhibit political, racial, or gender bias, or generate unintended and harmful outcomes. Therefore, AI agent discovery implementations must effectively filter out high-risk AI agents present within the unregulated Internet environment.\\

From a technical perspective, 'Agentic SafeSearch' does not constitute a discrete service but rather represents a sophisticated, filtered query mechanism to the NANDA Index API. This functionality is enabled through the enrichment of the AgentFacts data model with specific, verifiable trust and safety metadata. For instance, an agent's record within the NANDA Index may contain fields such as:

\begin{itemize}
    \item \textbf{trust\_certifications}: An array of cryptographically signed credentials from trusted third-party entities (e.g., 'kid-safe' certification from family safety organizations, or 'HIPAA-compliant' attestation for healthcare applications).
    \item \textbf{reputation\_scores}: A comprehensive set of scores from recognized agent auditing services, indicating reliability metrics, security posture assessments, and fairness evaluations.
    \item \textbf{content\_flags}: Self-declared or externally-verified flags indicating the nature of agent content classification (e.g., ["political", "financial\_advice", "adult\_content"]).
\end{itemize}

An 'Agentic SafeSearch' implementation subsequently translates safety policies (e.g., 'exclude political content, require kid-safe certification') into structured NANDA Index API queries such as:\\ /search?capability=...\&exclude\_flags=political\&requires\_cert=kid-safe-cert-v1. This approach transforms safety considerations from abstract objectives into verifiable, data-driven filtering mechanisms implemented at the infrastructure level.\newline

\section{Agent Visibility \& Control (AVC) and Zero Trust Agentic Access (ZTAA)}
AI agents operate within clearly defined business objectives and functional parameters, subject to explicit constraints governing permissible and prohibited actions. Agent providers maintain accountability for behaviors and actions executed by their AI agents. Conversely, agents may become targets of malicious entities or applications across Internet infrastructure, potentially being deceived into disclosing sensitive information or experiencing software compromise.\newline

\subsection{Zero Trust Agentic Access (ZTAA)}
Enterprise environments will increasingly witness the coexistence of personal AI agents and business-oriented agents. Both agent categories will access Internet resources and establish communication channels with remote agents—potentially spanning geographical distances of thousands of miles—to execute personal and business tasks concurrently. Contemporary corporate environments rely upon Zero Trust Network Access (ZTNA) frameworks to secure user interactions across Internet and web application infrastructure; a parallel paradigm is required for AI agents: Zero Trust Agentic Access (ZTAA).\\

\begin{figure}[H]
    \centering
    \includegraphics[height=2.5in,width=5in]{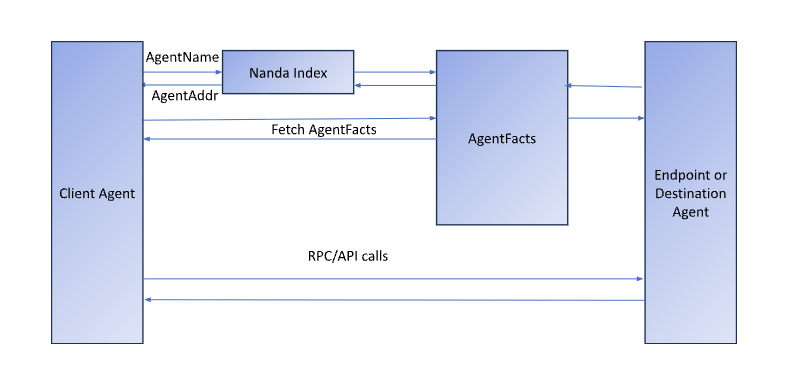}
    \caption{Protecting AI Agents}
    \label{fig:6}
\end{figure}

Zero Trust Agentic Access constitutes a security methodology that enforces a "never trust, always verify" approach for AI agent engagement with any external AI entities. This framework differs from the Zero Trust Network Access (ZTNA) model in that AI agents must not only verify identity and perform requisite additional authentication mechanisms (including Multi-Factor Authentication and agent security posture assessment), but also validate target AI agent capabilities, skill sets, geographical locations, and reputation metrics before establishing trust and communication channels with destination agents. Given that corporate AI agents may maintain access to both Internet networks and private applications, enforcing ZTAA protocols becomes critically important for both Internet-bound traffic and private access scenarios involving AI agents. Filtering and verification mechanisms utilizing NANDA's AgentFacts framework address these specific concerns, representing an indispensable component of agentic security architecture necessary to achieve Zero Trust Agentic Access (ZTAA).\\

The fundamental components within the NANDA architecture that facilitate ZTAA implementation include:

\begin{itemize}
    \item Implementation of cryptographically signed AgentFacts and credential validation logic that prevents agents from falsifying capabilities or engaging in impersonation attacks.

    \item Bilateral authentication and verification capabilities for source and destination agents against the NANDA registry, eliminating attack vectors for spoofing, reflection, and amplification attacks.

    \item Verification protocols with NANDA registry implementing multi-criteria filtering based on skills, geographical regions, safety classifications, categories, and reputation metrics, with skills, safety, and reputation maintained as third-party certifiable and verifiable attributes.
    
    \item Risk mitigation for "Newly Seen Agents" (NSAs): Analogous to the early developmental phase of the World Wide Web, the security landscape for agentic AI requires temporal maturation. During this transitional period, newly registered AI agents may possess limited verifiable information regarding their skills, capabilities, and reputation profiles. These "Newly Seen Agents" (NSAs) present significant risk factors, as consuming agents may lack sufficient contextual information for trustworthiness assessment. Development and implementation of effective risk mitigation strategies becomes critical to protect interacting agents from potential threats posed by NSAs.
\end{itemize}

\subsection{Agent Visibility and Control (AVC)}
As organizations progress toward commercialization of their AI agents, maintaining comprehensive records of agent-performed tasks and operational durations becomes essential for generating accurate invoicing and billing data.\newline

Business owners require critical oversight capabilities for agent operations and comprehensive traceability records of agent activities:

\begin{itemize}
    \item \textbf{Agent Identity Record Access:} Administrators must possess retrieval capabilities for complete identity information of any AI agent, encompassing agent nomenclature, Decentralized Identifier (DID), ownership attribution, and access permission specifications.

    \item \textbf{Performance History Access:} Administrators require review capabilities for agent historical task records, including completed task inventories, timestamp documentation, and task execution duration metrics.

    \item \textbf{Operational Control Authority:} Administrators must retain real-time authority to activate, pause, or terminate AI agent operations, ensuring appropriate oversight and intervention capabilities when necessary for safety, compliance, or ethical considerations.
\end{itemize}

\section{Governance and Compliance}
Drawing architectural parallels from Internet infrastructure frameworks, federal, local governmental, or business-specific compliance requirements may impose restrictions on AI agent transaction capabilities and inter-agent interactions within specific geographical regions. Many such restrictions can or should be enforced by computing infrastructure providers or security solution providers rather than AI agent providers. Nevertheless, agent providers must maintain awareness of such regulatory frameworks. \\

For instance, the US Department of Treasury mandates full compliance from US financial institutions with Office of Foreign Assets Control (OFAC)-governed sanctions regulations, ensuring non-engagement in trade or transaction activities that violate OFAC's country-based sanctions programs and avoiding trade or transaction activities with sanctions targets identified on OFAC's Specially Designated Nationals and Blocked Persons (SDN) list~\cite{ofac_2025}.\newline

The European Union's General Data Protection Regulation (GDPR) establishes comprehensive regulatory frameworks for protecting personal data of individuals within the EU and European Economic Area (EEA). It defines guidelines governing organizational data collection, processing, and storage of personal information. Organizations outside the EU remain subject to GDPR compliance when targeting or collecting data from EU-based individuals.\newline

Anticipated regulatory evolution will likely adapt similar governmental and regulatory frameworks to reflect governance requirements for AI agent Internet behavior and control mechanisms for agent-to-agent communications, including filtering capabilities and traceable event logging requirements.

\section{Conclusions and Forward Looking}
The emergence of agentic AI presents transformative opportunities across industrial sectors while simultaneously introducing significant risk vectors, including fraudulent activities, data breach incidents, and financial harm potential. For effective and secure agent functionality, users must maintain verification capabilities for both agent identity and claimed capability authenticity. As agentic AI infrastructure blurs traditional boundaries between personal and professional operational spheres, particularly within enterprise environments, a critical challenge emerges: constructing foundational frameworks that enforce IT policy compliance while preserving individual privacy rights. Addressing this fundamental tension represents an essential prerequisite for responsible agentic system deployment.\newline

\bibliographystyle{plain}
\bibliography{ref}

\end{document}